\documentclass{article}

\usepackage{graphicx}  
\usepackage{pstricks}
\usepackage{amsmath,amssymb} 

\newtheorem{thm}{Theorem}[subsection]
 \newtheorem{cor}[thm]{Corollary}
 
 \newtheorem{prop}[thm]{Proposition}
 \newtheorem{rem}[thm]{Remark}
 \newtheorem{proc}[thm]{Procedure}
 \numberwithin{equation}{subsection}

\title{Using the Eigenvalue Relaxation for Binary Least-Squares Estimation Problems}
\author{St\'ephane~Chr\'etien 
\thanks{M. Chr\'etien is with the Laboratoire de Math\'ematiques, UMR CNRS 6623 and Universit\'e de Franche Comt\'e, 
16 route de Gray, 25030 Besan{\c c}on Cedex, France. Email: stephane.chretien@math.univ-fcomte.fr}
and Franck Corset \thanks{M. Corset is with LabSAD Universit\'e Pierre Mendes France, 1251 Avenue centrale.
BP47 38040 Grenoble cedex 9, France. Email: franck.corset@upmf-grenoble.fr}}

\begin{document}

\maketitle

\begin{abstract}
The goal of this paper is to survey the properties of the eigenvalue relaxation for 
least squares binary problems. This relaxation is a convex program which is obtained as the 
Lagrangian dual of the original problem with an implicit compact constraint and as such, is a convex problem with polynomial 
time complexity. Moreover, as a main pratical advantage of this relaxation over the standard Semi-Definite Programming approach, 
several efficient bundle methods are available for this problem allowing to address problems of very large dimension. 
The necessary tools from convex analysis are recalled and shown at work for handling the problem of exactness of this relaxation. Two applications are described. The first one is the problem of binary image reconstruction and the second is the problem of 
multiuser detection in CDMA systems.  
\end{abstract}

\section{Introduction}
Several problems in engineering and in particular signal and image processing necessitate to estimate binary vectors 
corrupted by some noise and can be simply addressed using the least squares principle under binarity consraints. The resulting 
problem is a minimization of a quadratic form over $\{-1,1\}^n$, a problem which is known to be ${NP}$-Hard in general. One of the 
main approaches to relax this problem into a convex one is the Semi-Definite Programming relaxation which has been 
extensively used in classification, pattern recognition and communication systems. Some of the main achievements in the 
study of the SDP relaxation were obtained by Goemans and Williamson \cite{JACM95:Goemans} and \cite{OptMethSoft98:Nesterov}.
However, solving a SemiDefinite Program in practice relies on interior point methods which although enjoying nice theoretical 
convergence properties are limited to problems of size up to 500 $\times 500$. On the other hand, very pratically efficient 
bundle methods are available for the eigenvalue relaxation of the same binary quadratic optimization problems. 
We refer the reader to \cite{Lemarechal:Recent99} for a discussion of the pratical superiority of bundle methods for 
solving certain semi-definite programs such as the ones appearing in the present paper. Despite this empirical 
fact in favor of the eigenvalue relaxation, one of the main reasons most users prefer the SDP relaxation is that good primal binary solutions can be recovered using Goemans and Williamson's randomized algorithm. The main motivation of the present paper is 
to show how a solution of the SDP can be recovered from a solution of the eigenvalue relaxation. As a by product, a new geometric 
interpretation of the randomized algorithm is proposed. 

Penalized binary least squares estimation problems are problems of the form
\begin{equation}
\label{lsb}
\min_{x\in \mathbb R^n} \|y-Ax\|^2+\nu x^t P x
\hspace{.3cm}{\rm s.t. }\hspace{.3cm} x\in \{-1,1\}^n,
\end{equation}
where the vector $y\in \mathbb R^m$ is the observed data, the
matrix $A\in \mathbb R^{m\times n}$ represents the "filter", the
vector $x\in \mathbb R^n$ is the signal, or parameter vector, that
has to be estimated, and the term $\nu x^t Px$ is a penalization
term that can often be interpreted as an {\em a priori}
information in terms of Bayesian statistics.

This problem belongs to the larger class of minimization of
quadratic forms over binary vectors which is known to be ${NP}$-hard. Much work has been devoted to constructing Semi Definite Programming (SDP) based relaxations
for general quadratic binary problems. Semi-Definite programs are linear optimization problems over 
symmetric matrices with real coefficients and with the additional convex constraint of positive semidefiniteness; 
see for instance \cite{SIAM01:BenTal} or \cite{SIAM04:Boyd} for excellent introductions to convex programming and in particular SDP. 
SDP methods have already played an important role in various topics inside signal processing problems and 
we refer to \cite{MathPro03:Luo} for a nice survey on possible applications. 
A common feature of essentially all the existing relaxations is that they can be obtained using Lagrange duality 
which is a general methodology for obtaining lower bounds to hard minimization problems, as overviewed in
\cite{Pyth00:Lemarechal} and \cite{DAM02:Wolkowicz}. 

The goal of the paper is to survey what is known about another Lagrangian duality based relaxation, 
namely the eigenvalue relaxation, for this problem. This relaxation was first proposed by Delorme and Poljak 
\cite{MathProg93:Delorme} for the max-cut problem. 
See also the work of Poljak, Rendl and Wolkowics \cite{JGlobOpt95:Poljak} for more details. 
The main advantage of the eigenvalue relaxation over the SDP relaxation is that the eigenvalue relaxation can be solved 
much faster than the SDP relaxation, as reported for instance in 
\cite{Lemarechal:Recent99}, \cite{MathProg00:Oustry}, \cite{HandBookSDP00:Helmberg} and \cite{SIOPT00:Helmberg}.  
This remarkable computational tractability of the eigenvalue relaxation is the main motivation for writing 
this detailed survey. 

The content of the paper is as follows. The second section is devoted to a rapid presentation of the relaxation 
and its relationship with Lagrangian duality. We also recall a simple and well known certificate for exactness
of the relaxation, i.e. the fact that a globally optimal binary solution is obtained. 

The third section details the relationships between the Semi-Definite Relaxation and the eigenvalue relaxation. The 
main result of this section is the following: a solution of the SDP relaxation can be recovered from the solution of the 
eigenvalue relaxation. The case of inexact solutions to the eigenvalue relaxation is also studied. 

The forth section deals with the problem of recovering binary primal solutions from 
the dual scheme. We first give sufficient conditions under which strong duality holds and the eigenvectors of norm $n+1$ associated to the maximum 
eigenvalue at optimality are binary solutions. Next, in the case where strong duality does not hold, we show that Goemans and Williamson's randomized 
algorithm has a very natural meaning when viewed in terms of the optimal eigenspace associated to the maximal eigenvalue in the eigenvalue relaxation.  

In the last section, we propose simulation experiments
in the case of binary image denoising and CDMA Multiuser Detection problems. The first of these
problems has been previously approached by stochastic methods based
on Markov chains like simulated annealing and Metropolis Hastings
schemes; see for instance \cite{JRSS86:Besag} and the more recent work of
Gibbs \cite{Biometrika00:Gibbs}. The approach discussed here was presented in \cite{Compstat02:Chretien}. 
Recently a lot more problems have been addressed using the SDP relaxation in \cite{IEEEPAMI03:Keuchel}. 
The results obtained so far are quite
encouraging and the approach performs well on very dirty images. We prove hat strong duality holds for the immage 
denoising problem, thus recovering back the polynomial solvability result of Greig, Porteous and Seheult as a special case and
by a very different path.
Passing to the second problem, our Monte Carlo experiments show that the average computational effort 
for solving the eigenvalue relaxation as a function of the number of users grows slowlier than for 
the SDP relaxation with the standard implementations available with scilab.

{\bf Notations}. In the sequel we will use the following notations. The inner product on $\mathbb R^n$ is 
denoted by $\langle \cdot,\cdot \rangle$, the set of real symmetric matrices of order $n$ are denoted by $\mathbb S_n$.
The partial order $\succcurlyeq$ denotes the Loewner ordering, i.e. for $A$ and $B$ in $\mathbb S_n$,  
$A\succcurlyeq B$ means that $A-B$ is positive semidefinite. For a set $S$ in $\mathbb R^n$, ${\rm conv}(S)$ denotes the 
convex hull of $S$ and $\overline{S}$ denotes its closure. For a matrix $A$ in $\mathbb S_n$, $d(A)$ denotes its diagonal 
vector and for $a$ in $\mathbb R^n$, $D(a)$ denotes the diagonal matrix whose diagonal vector is $a$. If an 
equation number $\#$ corresponds to an optimization problem, then ${\rm opt}(\#)$ will denote the optimum value 
for this problem.

\section{The eigenvalue relaxation}

We first introduce the eigenvalue relaxation and at the same time,
we propose a quick refresher on Lagrangian duality, collecting all
the results that will play an essential role in the sequel. The
proofs of almost all the results presented here can be found in
\cite{ConvAnal93:Hiriart}.

\subsection{The Lagrangian dual and the eigenvalue relaxation}

The binary least-squares estimation problem is in fact
equivalent to the homogenized problem
\begin{equation}
{\max}_{x\in \mathbb R^{n+1}} -x^t \left[
\begin{array}{cc}
A^tA+\nu P & -A^ty \\
-y^tA & y^ty
\end{array}
\right] x  \text{ s.t. } x\in \{-1,1\}^{n+1}.
\hspace{2cm} {\rm (BLS)}
\nonumber
\end{equation}
Indeed, if we add the constraint $x_{n+1}=1$ in (BLS), we
obtain exactly the binary least squares problem. Now, if $x^*$ is a solution of (BLS), then $-x^*$ is again a solution of
of (BLS), thus adding the constraint $x_{n+1}=1$ is in fact
redundant, which proves the claimed equivalence. Set
\begin{equation}
M=\left[
\begin{array}{cc}
A^tA+\nu P & -A^ty \\
-y^tA & y^ty
\end{array}
\right].
\nonumber
\end{equation}

Notice further that the constraint $x_i\in \{-1,1\}$ is equivalent
to $x_i^2=1$ for all $i=1,\cdots,n+1$. Thus, to problem
(BLS), we can associate the Lagrangian function
\begin{equation}
\begin{array}{rl}
L(x,u) & =-x^tMx+\sum_{i=1}^{n+1} u_i(x_i^2-1)\\
       & =x^t(D(u)-M)x-u^te.
\end{array}
\nonumber
\end{equation}

Now we can add to the problem the implicit spherical constraint
$$\mathcal S_{n+1}=\{ x\in \mathbb R^{n+1} \mid x^tx=n+1 \},$$
which is redundant with the binary constraints. Then, optimizing
over this sphere, we obtain the Lagrangian dual function, i.e.
\begin{equation}
\begin{array}{rl}
\theta(u) & ={\max}_{x\in \mathcal S_{n+1}} x^t(D(u)-M)x-u^te \\
& =\max_{x\in \mathcal S_{n+1}} x^t(D(u)-M)x-\frac{u^te}{n+1}x^tx \\
& =\max_{x\in \mathcal S_{n+1}}
x^t\Big(D(u)-M-\frac{u^te}{n+1}I\Big)x
\end{array}
\nonumber
\end{equation}
which, using Raleigh-Ritz variational formulation of the largest
eigenvalue of symmetric matrices, can be written
\begin{equation}
\theta(u)=(n+1)\lambda_{\max}\Big(D(u)-M-\frac{u^te}{n+1}I\Big).
\end{equation}

Finally, the dual problem, i.e. the eigenvalue relaxation, is
given by
\begin{equation}
\label{eigrlx}
\min_{u\in \mathbb R^{n+1}} \theta(u).
\end{equation}

\subsection{Properties of the dual relaxation}

\subsubsection{Convexity}

It is important to notice first that the dual function $\theta(u)$
is convex, since it is the maximum over a family parametrized by $x\in \mathcal
S_{n+1}$ of linear functions in the variable $u$.

\subsubsection{Weak duality}

The main classical property of the Lagrangian dual is weak
duality, i.e.
\begin{equation}
\min_{u\in \mathbb R^{n+1}} \theta(u) \geq {\rm opt}\rm{(BLS)},
\nonumber
\end{equation}
where ${\rm opt}$ denotes the optimal value. 

This property explains in part why Lagrange duality is used : it
provides a bound on the primal optimal value. When equality holds in the
weak duality property, we say that strong duality holds.
Sometimes, like in the case of the Max-Cut problem, the bound can
be proved to be proportional to the optimal original value. More
precisely, Goemans and Williamson proved that the optimum value of
the eigenvalue relaxation (in fact the equivalent SDP formulation; see 
the original paper and Section \ref{eqSDP} below)
is greater than or equal to the optimal original value (this is
just weak duality), which itself is always greater than or equal
to .876 times the eigenvalue relaxation's optimal value. A quite
similar but less tight bound, proved by Nesterov applies directly
to the present problem. We will recall this bound in section
\ref{GWN} below.

\subsubsection{Existence of dual solutions}

It is well known that there exists an optimal dual solution. This
was proved by Poljak and Wolkowicz in \cite{MathOR95:Poljak}. The
proof given here is more direct.

\begin{prop}
The dual function admits a minimizer.
\end{prop}
{\bf Proof}. 
Let $\theta^*=\inf_{u\in \mathbb R^{n+1}}\theta(u)$. Make the
change of variable $v=u-\frac1{n+1} \sum_{i=1}^{n+1} u_i$, i.e.
define
\begin{equation}
\eta(v)=(n+1)\lambda_{\max}(D(v)-M)=\theta(u).
\nonumber
\end{equation}
We now have the property that $\sum_{i=1}^{n+1}v_i=0$. We prove
that $\eta$ is coercive. Take any sequence $(v^k)_{k\in \mathbb
N}$ with $\|v_k\|\rightarrow +\infty$ as $k\rightarrow +\infty$.
We can assume that $v^{k}_i \rightarrow +\infty$ for some $i$
because otherwise, the fact that $\|v_k\|\rightarrow +\infty$
implies that there must exists a sequence $(v_j^k)_{k\in\mathbb
N}$ with $v^{k}_j \rightarrow -\infty$ and the fact that
$\sum_{i=1}^{n+1}v_i=0$ gives a contradiction. Now, the Gershgorin
circle around the diagonal element $M_{i,i}+v^k_i$ has a constant
radius, say $r$ and its center goes to $+\infty$. Since
$|M_{i,i}+v^k_i-\lambda_{\max}(D(v)-M)|$, this implies that
$\lambda_{\max}(D(v)-M)\rightarrow +\infty$. Thus $\eta$ is
coercive and since it is continuous, it admits a minimizer that we 
will denote by $v^*$. Now, for all $u\in \mathbb R$, $v=u-\frac1{n+1} \sum_{i=1}^{n+1} u_i$, we
have
$$\theta^*\leq \theta(v^*)$$
But, on the other hand, $\theta(v^*) =\eta(v^*) \leq
\eta(v)=\theta(v)=\theta(u)$. Therefore,
$$\theta(v^*) \leq \theta^*$$
and the proof is complete.
\hfill $\Box$

\subsubsection{Subdifferential's description and exactness criterion}
The subdifferential $\partial \theta(u)$ of the eigenvalue
relaxation has been much studied. Recall that for any convex
function $f:\mathbb R^m\mapsto \mathbb R$, the subdifferential
$\partial f(u)$ is defined by
$$\partial f(u)=\Big\{ g\in \mathbb R^m \mid f(u^\prime)\geq f(u)+g^t(u^\prime-u)\Big\}.$$
The analysis of $\partial \theta(u)$ is based on the following
general theorem.
\begin{thm}\label{eigsubd}{\rm \cite{ConvAnal93:Hiriart}}
Let $A: \mathbb R^m \mapsto \mathbb S_n$ be an affine operator
defined by $A(u)=\mathcal A u+B$ for some linear operator
$\mathcal A: \mathbb R^m \mapsto \mathbb S_n$ and some matrix
$B\in \mathbb S_n$. Then, we have
$$\partial (\lambda_{\max}\circ A)(u))=\mathcal A^* \partial \lambda_{\max}(A(u))$$
with
$$
\begin{array}{c}
\partial \lambda_{\max}(X)=\\ E_{\max} \Big\{Z\in \mathbb S_{r_{\max}}\mid Z\succcurlyeq 0 
\text{ and } {\rm trace}(Z)=1 \Big\}E_{\max}^t 
\end{array}
$$
where $\mathcal A^*$ is the adjoint of $\mathcal A$, $r_{\max}$
denotes the multiplicity of $\lambda_{\max}$ at $X\in \mathbb S_n$
and $E_{\max}$ is a matrix whose columns form any orthonormal
basis of the eigenspace of $X$ associated to $\lambda_{\max}$.
\end{thm}
Now, if we set $A(u)=D(u)-M-\frac{u^te}{n+1}I$, we get $B=-M$,
$\mathcal Au=D(u)-\frac{u^te}{n+1}I$ and $\mathcal
A^*X=d(X)-\frac1{n+1}{\rm trace}(X)e$. For $d\in\mathbb N$, let
$\mathcal Z_d$ be defined by
$$\mathcal Z_d=\Big\{Z\in \mathbb S_{d}\mid Z\succcurlyeq 0 \text{
and } {\rm trace}(Z)=1 \Big\}.$$ Using the previous theorem, we
obtain
\begin{cor}
\label{subdtheta}
The subdifferential $\partial \theta(u)$ of the dual function
$\theta$ is given by
$$\partial \theta(u)=(n+1)d(E_{\max} \mathcal Z E_{\max}^t)
- {\rm trace}(E_{\max} \mathcal Z E_{\max}^t)e
$$
\end{cor}
Following Oustry \cite{MathProg00:Oustry}, the formula for $\partial
\lambda_{\max}(X)$ in theorem \ref{eigsubd} is proved by showing
that the maximum eigenvalue function $\lambda_{\max}(X)$ on
$\mathbb S_n$ is nothing but the support function
$\sigma_{\mathcal Z_n}(X)$ of $\mathcal Z_n$, defined by
\begin{equation}
\label{support} \sigma_{\mathcal Z_n}(X)=\sup_{Z\in \mathcal Z_n}
\langle X,Z\rangle
\nonumber
\end{equation}
with the scalar product defined by $\langle X,Z\rangle={\rm
trace}(X,Z)$. By definition, the face $F_{\mathcal Z_n}(X)$ of
$\mathcal Z_n$ exposed by $X$ is the set of maximizers in
(\ref{support}), i.e.
$$ F_{\mathcal Z_n}(X)=\Big\{Z\in\mathcal Z_n \mid \lambda_{\max}(X)=\langle X,Z\rangle \Big\}.$$
Knowing that the subdifferential of a support function of a set is
exactly the exposed face of this set, we finally get
$$\partial \lambda_{\max}(X)=\Big\{Z\in\mathcal Z_n \mid \lambda_{\max}(X)=\langle X,Z\rangle \Big\} $$
the formula follows after some linear algebra. 

There is a different path to the subdifferential's formula, which
is perhaps more {\em a propos} in the context of duality: it is
proved in \cite[Chapter XII]{ConvAnal93:Hiriart} that
\begin{equation}
\label{subd2}
\partial \theta(u)=\overline{{\rm conv}}\Big\{
(x_1^2-1,\cdots,x_{n+1}^2-1)^t \mid L(x,u)=\theta(u)\Big\},
\end{equation}
where $\overline{\rm conv}$ denotes the closure of the
convex hull. This fact is in fact true for general continuous
constrained problems in the case where the underlying space is
compact (for example) \footnote{which is the case here since we
optimize over the sphere $\mathcal S_{n+1}$} and the associated
technical condition is called the {\em filling property}. The
following proposition provides a useful sufficient condition for
proving that the relaxation is exact, i.e. strong duality applies.

\begin{prop}
\label{multone} Let $u^*$ be a minimizer of the dual eigenvalue
relaxation. Then, if $\lambda_{\max}(A(u^*))$ has multiplicity
one, then
$$
\min_{u\in \mathbb R^{n+1}} \theta(u^*) = {\rm opt}{\rm (BLS)}
$$
and any eigenvector $x$ of $A(u^*)$ whose squared norm is $n+1$ is
a binary solution of (BLS).
\end{prop}
The proof is a direct consequence of
\cite[Theorem XII.2.3.4.]{ConvAnal93:Hiriart}. We provide a specialized
proof here because it is short and instructive.

{\bf Proof}. 
Since the multiplicity of $\lambda_{\max}(A(u^*))$ is one, the
subdifferential of $\lambda_{\max}\circ A$ at $u^*$ is a single
vector. Thus, $\theta$ is differentiable at $u^*$ and its gradient
is simply
$$\nabla \theta(u^*)=({x_1^*}^2-1,\cdots,{x_{n+1}^*}^2-1)^t$$
for any $x^*$ in $\mathcal S_{n+1}$ such that
$\theta(u^*)=L(x^*,u^*)$. Since, $u^*$ minimizes $\theta$, we must have $\nabla \theta(u^*)=0$.
This implies that ${x_i^*}^2=1$ for all $i=1,\cdots,n+1$. Thus, using weak duality
$$ {\rm opt}({\rm BLS})\leq \theta(u^*)={x^*}^t(-M)x^*\leq
{\rm opt}({\rm BLS})$$ which proves that $x^*$ solves the original
problem ({\rm BLS}).
\hfill $\Box$

We now have a nice criterion for deciding whether our
relaxation was exact and if so, we also know how
to recover a binary solution from an optimal eigenvector. This approach works for any
quadratic binary problem and is extensively used for approximating
combinatorial problems. However, the question remains on what to
do when the relaxation is not exact, i.e. when the multiplicity at
the optimum is greater than one. The next two sections will help
answer this crucial question.

\section{From eigenvectors to SDP solutions}
\label{eqSDP}
The purpose of the next two sections is to describe how to recover
primal binary solutions from the eigenvector solutions of the dual 
eigenvalue problem. It was first shown that good binary
solution can be generated at random using the SDP solution by
Goemans and Williamson \cite{JACM95:Goemans} in the case of the
Max-Cut problem in graph theory. Their results were then extended
by Nesterov to the case of indefinite quadratic binary programming
\cite{OptMethSoft98:Nesterov}. Those results allowed to conclude that both
eigenvalue and SDP relaxations are in a certain precise sense very
efficient. However, both relaxations are not equivalent from the
computational point of view. Recall that one of the main motivations for using the
eigenvalue relaxation is its manageable practical
complexity which is often favorable compared to the one of solving
the SDP relaxation. But what is not clear is how to generate good
(primal) binary solutions in average with the eigenvalue
relaxation only ? The first natural approach to this question is
of course to try and recover an optimal SDP solution from the
eigenvalue relaxation. Thus, we devote this section to this
problem. It can be solved as follows : an appropriate convex
combination of rank one matrices obtained from a set of optimal
eigenvectors is shown to be a solution we are looking for. Our
approach simplifies the presentation of
\cite{SIOPT95:Poljak}. The adaptation of the randomized
algorithm of Goemans and Williamson and the associated bound
established by Nesterov will be discussed in the next section.

\subsection{The SDP relaxation}
\label{SDP}
In order to obtain the Semi-Definite Programming (SDP) relaxation of the the homogenized
problem ({\rm BLS}), we begin with the following equivalence relating our problem to a problem on 
symmetric matrices. We have \footnote{Here, we use the fact that $x^tMx={\rm trace}(x^tMx)={\rm trace}(Mxx^t)$}
$${\rm opt}({\rm BLS})=\max_{x\in\mathbb R^{n+1}} {\rm trace}(-Mxx^t) \text{ s.t. } d(xx^t)=e. $$
This last problem is itself equivalent to
$$ \max_{X\in\mathbb S_{n+1}} {\rm trace}(-MX) \text{ s.t. } d(X)=e, \:\: X\succcurlyeq 0,\:\: {\rm rank} X=1.$$
This problem being nonconvex, we drop the rank constraint and
obtain the following SDP (convex) relaxation
\begin{equation}
\max_{X\in\mathbb S_{n+1}} {\rm trace}(-MX)\text{ s.t. } d(X)=e,
\:\: X\succcurlyeq 0 \hspace{1cm}
\hfill {\rm (SDP)}
\nonumber
\end{equation}
whose value is obviously greater than or equal to ${\rm
val({\rm BLS})}$.

An important result of Pataki \cite[Theorem 2.1]{Pataki:MOR98} gives a bound on the rank of solutions to Semi-Definite Programs. In the case of 
our Semi-Definite relaxation, this theorem implies that the rank $r^*$ of an optimal matrix $X^*$ satisfies $\frac12 r^* (r^*+1)\leq n$. 

\subsection{SDP versus maximal eigenvalue : theoretical
equivalence} It follows from the subdifferential's formula given in Corollary \ref{subdtheta} that at
any minimizer $u^*$, we have
$$
\begin{array}{c}
0\in \partial \theta(u^*)= \\
(n+1)d(E_{\max}^* \mathcal Z_{r^*_{\max}}
{E_{\max}^*}^t) - {\rm trace}(E_{\max}^* \mathcal Z_{r^*_{\max}}
{E_{\max}^*}^t)e.
\end{array}
$$
Suppose we have in hand a matrix $Z^*\in \mathcal Z_{r^*_{\max}}$ such that
\begin{equation}
\label{optZ} 0=(n+1)d(E_{\max}^* Z^* {E_{\max}^*}^t) -
{\rm trace}(E_{\max}^* Z^* {E_{\max}^*}^t)e.
\end{equation}
It appears that a good guess for a candidate solution $X^*$ to the
SDP relaxation in the general case is
$$X^*=(n+1)E_{\max}^* Z^* {E_{\max}^*}^t.$$
We just need to check the details to see how it works. This result was initially proved in 
\cite{SIOPT95:Poljak} but the proof given here is more direct. 
\begin{thm}\cite{SIOPT95:Poljak}
\label{equiv} Let $u^*$ be the optimal solution of the eigenvalue
relaxation let $E_{\max}$ be a matrix whose columns for an
orthonormal basis of the eigenspace associated to
$\lambda_{\max}(A(u^*))$ and let $Z^*$ be as in (\ref{optZ}). Then
the matrix $X^*=(n+1)E_{\max}^* Z^* {E_{\max}^*}^t$ is an optimal
solution of the SDP relaxation.
\end{thm}

\begin{rem}
We would like to underline at this point that a more elegant proof
of the theorem could be obtained using conic duality but we
preferred to keep on with elementary arguments since this is
possible in the present context.
\end{rem}
{\bf Proof}. 
Compute the eigenvalue/eigenvector decomposition $Z^*=U\Delta
U^t$, set $F=E_{\max}^*U$, $\delta=d(\Delta)$, let $r$ be the
multiplicity of $A(u^*)$ and let $f_1,\cdots,f_{r}$ denote the
columns of $F$. Recall that from the definition of $Z^*$, we have
$\sum_{j=1}^r \delta_j=1$. Then, we get
$$
0=d(F\Delta  F^t) -\frac1{n+1} {\rm trace}(F \Delta F^t)e.
$$
Thus,
$$
\begin{array}{c}
{\rm trace}\Big((D(u^*)-\frac1{n+1}(u^*)^teI)F\Delta F^t\Big) =\\
(u^*)^td(F\Delta F^t)-(u^*)^t\frac1{n+1}{\rm trace}(F\Delta
F^t)e =0.
\end{array}
$$
Using this fact, we obtain
$$
\begin{array}{l}
{\rm trace} (-MX^*) \\ 
=(n+1){\rm trace}\Big((-M+D(u^*)-\frac1{n+1}(u^*)^te I)F\Delta F^t\Big) \\
= (n+1){\rm trace} (A(u^*)F\Delta F^t)\\
= (n+1){\rm trace} (A(u^*)\sum_{j=1}^r\delta_j f_jf_j^t)\\
= (n+1) \sum_{j=1}^r \delta_j f_j^tA(u^*) f_j \\
= (n+1) \sum_{j=1}^r \delta_j \lambda_{\max}(A(u^*)) \\
= (n+1)\lambda_{\max}(A(u^*)), 
\end{array}
$$
since $\sum_{j=1}^r \delta_j=1$. 
Thus, the optimal value of the SDP is greater than or equal to the
optimal value of the eigenvalue relaxation. On the other hand, it 
is well known that the optimal value of the eigenvalue
relaxation is greater than or equal to the one of the SDP
relaxation. We provide a proof here for the sake of completeness.
Let $X^{**}$ be an optimal solution to the SDP relaxation. Now,
for all $u$ in $\mathbb R^{n+1}$, we have
$$
{\rm trace} \Big(X^{**}(D(u)-\frac{e^tu}{n+1}I)\Big)=0
$$
by using the fact that $D(X^{**})=e$. Now, compute the
eigenvalue/eigenvector decomposition
$-M+D(u)-\frac{e^tu}{n+1}I=\sum_{i=1}^{n+1}\lambda_i v_iv_i^t$ and
let $\lambda_{\max}$ be the greatest of these eigenvalues. Then,
$$
\begin{array}{rl}
{\rm trace}(-MX^{**}) & ={\rm
trace}\Big(X^{**}(-M+D(u)-\frac{e^tu}{n+1}I)\Big) \\
&=\sum_{i=1}^{n+1} \lambda_i v_i^t X^{**}v_i \\
&\leq \lambda_{\max} \sum_{i=1}^{n+1} v_i^t X^{**}v_i \\
&= \lambda_{\max} {\rm trace}(X^{**}\sum_{i=1}^{n+1}v_iv_i^t) \\
&=\lambda_{\max} {\rm trace}(X^{**}I) \\
&=(n+1) \lambda_{\max}
\end{array}
$$
Since this is true for all $u$, we obtain that the eigenvalue
relaxation majorates the SDP relaxation. Thus, both optimal values
are equal and this completes the proof of the proposition.
\hfill $\Box$
\subsection{SDP versus maximal eigenvalue: practical
implementation} \label{pract} Of course, it can be hard to find a
matrix $Z^*\in \mathcal Z_{r^*_{\max}}S$ that works. We will now try to overcome
this problem. We first have to specify how the subgradients are
obtained in practice. At each point $u\in \mathbb R^{n+1}$, choose
an eigenvector $x$ of squared norm equal to $n+1$ associated to
$\lambda_{\max}(A(u))$. Then, using the alternative representation
of the subdifferential (\ref{subd2}), a subgradient of $\theta$ at
$u$ is obtained by setting $g=[{x_1}^2-1,\ldots,{x_{n+1}}^2-1]^t$.
Assume that we have a set of subgradients
$g_j=[{x_1^j}^2-1,\ldots,{x_{n+1}^j}^2-1]^t\in
\partial \theta(u^j)$ for some $u^j$, $j=1,\ldots,p$ and such that
\begin{equation}
\label{nullconvexcomb} \|0-\sum_{i=1}^{p}\alpha_j g_j\|\leq
\epsilon, \hspace{1cm} {\rm (\epsilon OPT)}
\nonumber
\end{equation}
for some nonnegative $\alpha_j$'s with $\sum_{j=1}^{p}\alpha_j=1$.
This can be performed for $\epsilon$ as small as we want by using
a bundle method. Such a method will construct in a finite number
of iterations, say $k$, an iterate $u^k$ and a family of $u^j$'s
with the desired property, all of them lying in a small
neighborhood of $u^k$. This is one very nice feature of the bundle
mechanism which is extensively described in \cite[Volume
II]{ConvAnal93:Hiriart}. Moreover, it is a well known fact, called
Caratheodory's theorem, that only $p=n+2$ subgradients are
sufficient in the expression ($\epsilon$OPT).

Set
$$X^*_{\epsilon}=\sum_{j=1}^{p}\alpha_j x^j{x^j}^t.$$
Then, we have the following result.
\begin{prop}
For any $\epsilon>0$, the matrix $X^*_{\epsilon}$ defined above
satisfies
$${\rm trace}(MX^*_{\epsilon})\leq \min_{u\in \mathbb R^{n+1}} \theta(u)-\mathcal O(\epsilon).$$
\end{prop}
{\bf Proof}. 
Let $u^*$ be any minimizer of $\theta$. Then, for each
$j=1,\ldots,p$, we have by the definition of the subdifferential
$$\theta(u^*)\geq \theta(u^j)+g_j^t(u^*-u^j).$$
But $\theta(u^j)$ is given by
$$\theta (u^j)={x^j}^t\Big(D(u^j)-M-\frac{e^tu^j}{n+1}I\Big)x^j.$$
On the other hand, since ${x^j}^tx^j=n+1$,
$$\begin{array}{l}
{x^j}^t \Big(D(u^j)-M-\frac{e^tu^j}{n+1}I\Big)x^j \\ 
={x^j}^tMx^j+\sum_{i=1}^{n+1}u_i{x_i^j}^2-\sum_{i=1}^{n+1}u_i\\
={x^j}^tMx^j+\sum_{i=1}^{n+1}u_i({x_i^j}^2-1) \\
={x^j}^tMx^j+g_j^tu^j.
\end{array}
$$
Thus, we obtain
$$\theta(u^*)\geq {\rm trace}(M{x^j}^tx^j)+g_j^tu^*$$
which implies, after multiplying by $\alpha_j$ and summing over
$j=1,\ldots,p$
$$\theta(u^*)\geq {\rm trace}(MX^*_{\epsilon})+(\sum_{j=1}^{p}\alpha_jg_j)^tu^*.$$
Using Cauchy-Schwartz inequality, this gives
$$\theta(u^*)\geq {\rm trace}(MX^*_{\epsilon})+\epsilon \|u^*\|.$$
Since the eigenvalue and the SDP relaxation have equal optimal
values, we finally obtain
$${\rm opt(SDP)} \geq {\rm trace}(MX^*_{\epsilon})+\epsilon \|u^*\|$$
which implies the desired result.
\hfill $\Box$

\subsection{Comments} It is a common idea that the SDP relaxation
contains more information than the eigenvalue relaxation. We hope
that the results of this section managed to convince the reader
that this is in fact not the case and a good approximate solution
can be recovered quite easily using subgradient information at the optimum.

\section{Recovering primal binary solutions}
We now are in position to answer our main question of how to
recover a satisfactory although sometimes suboptimal primal binary solution.
In the first part of this section, we show that optimal binary solutions can actually be exactly recovered using 
the eigenvalue relaxation, i.e. strong duality holds, under some simple conditions. 
Then, in the case where the problem does not satisfy these necessary conditions for strong duality, 
we develop a randomized algorithm based on the optimal 
eigenspace of the maximum eigenvalue dual function and show that this procedure 
is equivalent to Goemans and Williamson's randomized algorithm for Max-Cut. This provides 
a new interpretation of Goemans and Williamson's procedure.  

\subsection{A sufficient conditions for strong duality}

We have the following theorem. 
\begin{thm}
\label{strong}
For almost all $A$ in the sense of the Lebesgue measure, 
such that $A^tA+\nu P$ is componentwise negative outside the diagonal. Then the eigenvalue relaxation is exact, i.e. strong duality holds.  
\end{thm}

{\bf Proof}. Fix $u\in \mathbb R^{n+1}$. Let $u_1^n$ be the vector of the first $n$ components of $u$. 
The fact that $A^tA+\nu P$ is componentwise negative outside the diagonal implies that $-A^tA-\nu P+D(u_1^n)-\min(u_1^n)I$ is componentwise 
positive. Thus, the Perron-Frobenius theorem implies that the maximum eigenvalue of 
$-A^tA-\nu P+D(u_1^n)-\min(u_1^n)I$ has multiplicity one. From this, we deduce that 
the maximum eigenvalue of $-A^tA-\nu P+D(u_1^n)$ also has multiplicity one. Let $V_{u_1^n}D_{u_1^n}V_{u_1^n}^t$ be an eigenvalue 
decomposition of $A^tA+\nu P+D(u_1^n)$, where we used the subscript $u_1^n$ in order to remember that whatever the chosen decomposition, it 
is a nonlinear and non necessarily continuous function of $u$. Moreover, since the maximum eigenvalue has multiplicity 
one, Corollary 4 in \cite{Meyer:SIAMNA88} says that it is possible to choose the eigenvector associated to the maximum eigenvalue 
as a continuously differentiable function of $u_1^n$. We will denote by $v_{u_1^n}^{\max}$ this eigenvector. 
Using this parametrization, the matrix 
\begin{equation}
-M+D(u)=-\left[
\begin{array}{cc}
A^tA +\nu P & -A^ty \\
-y^tA & y^ty
\end{array}
\right]+D(u)
\nonumber
\end{equation}
can be rewritten as 
\begin{equation}
-M+D(u)=\left[
\begin{array}{cc}
V_{u_1^n} & 0 \\
0 & 1
\end{array}
\right]
\left[
\begin{array}{cc}
D_{u_1^n} & -V_{u_1^n}^tA^ty \\
-y^tA V_{u_1^n} & y^ty+u_{n+1}
\end{array}
\right]
\left[
\begin{array}{cc}
V_{u_1^n} & 0 \\
0 & 1
\end{array}
\right]^t,
\nonumber
\end{equation}
where all dimensions can easily be guessed from the previous knowledge on the involved submatrices. 

Let $\mathcal V$ be the codimension one differentiable submanifold defined by $$\mathcal V=\{ (A,u) \in \mathbb R^{m\times n} \times \mathbb R^{n+1}\mid y^tAv_{u_1^n}^{\max}=0 \}.$$ 
Let $\mathcal W$ be the optimal set defined by $$\mathcal W=\{ (A,u) \in \mathbb R^{m\times n} \times \mathbb R^{n+1}\mid 0\in \partial \theta(u)\}.$$
Due to the representation 
\begin{equation}
\begin{array}{rc}
\partial \theta(u) & =\{V_{\max} Z V_{\max}^t \mid A \in \mathbb R^{m\times n}, u\in \mathbb R^{n+1},
\:V\in \mathbb R^{(n+1)\times r_{\max}}, \: Z \in \mathbb S_{r_{\max}}, \: Z\succeq 0, \\  &
\\  &
(-M+D(u))V_{\max}=\lambda V_{\max}, V^t_{\max} V_{\max}=I, \: {\rm trace}(Z)=1 \},
\end{array}
\nonumber
\end{equation}
the set $\mathcal W$ is the projection onto the cartesian product $\{ (A,u) \in \mathbb R^{m\times n}\times \mathbb R^{n+1}\}$ of the set $\cup_{r=1}^R 
\tilde{W}_r$ where $R$ is the upper bound of Pataki  (see Section \ref{SDP}) on the optimal rank of the SDP 
relaxation\footnote{which also holds for the eigenvalue relaxation due to the complete equivalence between these two problems} 
(here $R\leq \sqrt{2n}$ for $n$ large) and where $\tilde{W}_r$ is the set
\begin{equation}
\label{optmanifold}
\begin{array}{rc}
\tilde{W}_r & =\{(A,u,V,\lambda,Z) \mid A \in \mathbb R^{m\times n}, u\in \mathbb R^{n+1},
\: V\in \mathbb R^{(n+1)\times r}, \: Z \in \mathbb S_{r},\: (-M+D(u))V=\lambda V, \\ & 
\\  & V^t V=I, \: {\rm trace}(Z)=1,\: (n+1) d(V Z V^t)+{\rm trace}(V Z V^t) e=0 \},
\end{array}
\nonumber
\end{equation}
whose intersection with $\{(A,u,V,\lambda,Z) \mid A \in \mathbb R^{m\times n}, u\in \mathbb R^{n+1},
\: V\in \mathbb R^{n\times r}, \: Z \in \mathbb S_{r}, \: Z\succeq 0\}$ corresponds to the parameter 
set allowing for zero to belong to the subdifferential of the dual function $\theta$ in the case where $u=\lambda_{\max}(-M+D(u))$. Now, since 
the constraint $(-M+D(u))V=\lambda V$ is described by $(N+1)r$ equations, $V^t V=I$ by $r\frac{(r+1)}2$ equations, ${\rm trace}(Z)=1$ 
by one equation and $(n+1) d(V Z V^t)+{\rm trace}(V Z V^t) e=0$, the dimension of $\tilde{W}_r$ is greater than or equal to 
$m\times n+(n+1)+(n+1)\times r+1+r\times \frac{(r+1)}2-(n+1)\times r-r\times \frac{(r+1)}2-1+(n+1)=m\times n$. Furthermore, notice that
since the eigenvalues are continuous fonctions of the entries of $-M+D(u)$,  
the subset of $\cup_{r=1}^R \tilde{W}_r$ for which $u=\lambda_{\max}(-M+D(u))$ is open in the topology induced by the ambiant space. Therefore 
its projection set onto the cartesian product $\{ (A,u) \in \mathbb R^{m\times n}\times \mathbb R^{n+1}\}$ is of dimension at least $m\times n$
which garantees that the projection onto the $A$-space $\{ A \in \mathbb R^{m\times n}\}$ of its intersection with $\mathcal V$ is a set of null 
Lebesgue measure. And thus, for almost all $A$, such that $A^tA+\nu P$ is componentwise negative outside the diagonal, $y^tAv_{u_1^n}^{\max}\neq 0$. 

Using this result, Theorem A about the 
interlacing property of the eigenvalues for arrow matrices in the Appendix implies that the maximum 
eigenvalue of $M+D(u)$ is greater than the maximum diagonal element of $D_{u_1^n}$ which nothing by $\lambda_{\max}(-(A^t A+\nu P)+D(u_1^n))$ 
and all $n$ other eigenvalues are less than $\lambda_{\max}(-(A^t A+\nu P)+D(u_1^n))$. This implies that for allmost all $A$, the maximum 
eigenvalue of $M+D(u)$ has multiplicity one at the optimum, which implies that $\theta$ is differentiable at the optimum. 
Therefore, using Proposition \ref{multone} we obtain that strong duality holds for allmost all $A$ such that $A^tA+\nu P$ is componentwise negative
outside the diagonal.   
\hfill$\Box$

\subsection{When strong duality fails: the randomized algorithm}

We start this section with some recalls on Goemans and Williamson's algorithm and 
Nesterov's bound. 
\subsubsection{Goemans and Williamson's algorithm and Nesterov's bound}
\label{GWN}
The method relies on the Cholesky factorization of the optimal
solution $X^*$ of the SDP relaxation,
$$X^*=V^tV.$$
From Theorem \ref{equiv} we see that $V\in \mathbb R^{(n+1)\times
r_{\max}}$ where $r_{\max}$ is the multiplicity of
$\lambda_{\max}(A(u^*))$ at the chosen corresponding solution
$u^*$ of the eigenvalue relaxation. This factorization is
important, since it allows to write $X_{ij}^*=v_i^tv_j$ where
$v_i$ is the transpose of $i^{th}$ row vector of $V$. Let $\xi$ be
a random variable with uniform distribution on the unit sphere in
$\mathbb R^{r_{\max}}$.
\begin{proc}
\label{SDPGW}
{\bf (Goemans and Williamson's algorithm)}

1. Find the Cholesky factorization $X^*=V^tV$.
 
Let $\zeta$ be a random vector with uniform distribution on the
unit sphere of $\mathcal S(0,1)$. The random cut is defined by
$$
Z={\rm sign} \Big( V^t\zeta \Big).
$$
where the sign function is defined coordinate-wise.  

2. Draw $n$ samples from $Z$, say $z^1$, \ldots, $z^n$ and choose
the sample giving the best value of the objective function ${z}^tMz$.

\end{proc}
The key result is that, in average, the vector $Z$ gives a good binary solution to the
original problem. Since the best sample will have greater cut value than the average with 
overwhelming probability, the above procedure should work well. This is made precise by Nesterov's theorem.
\begin{thm}[Nesterov]
Define
$$f^*=\max_{x \in \mathbb R^{n+1}} x^tMx \text{ s.t. } x\in \{-1,1\}^{n+1} $$
and
$$f_*=\min_{x \in \mathbb R^{n+1}} x^tMx \text{ s.t. } x\in \{-1,1\}^{n+1} $$
then, we have
$$
\frac{f^*-E[z^tMz]}{f^*-f_*}\leq \frac2{\pi}.
$$
\end{thm}
This result is remarkable despite the fact that the bound
$\frac2{\pi}$ is rather large. An important issue for future research is to study such type of
bounds for particular subclasses of problems in hope of improving
Nesterov's result.

\subsubsection{The eigenvector viewpoint}
The main drawback of the former presentation is that using the
uniform variable $\xi$ is quite hard to motivate from an
optimization viewpoint. Let us take a slightly different
perspective. Assume that we have a solution $u^*$ of the
eigenvalue relaxation. As before, let $E_{\max}$ be a matrix whose
columns form an orthonormal bases of the eigenspace associated to
$\lambda_{\max}(A(u^*))$. Moreover, we may require that
\begin{equation}
\label{eigopt}
0=\mathcal A^*(E_{\max}\Delta E_{\max}^t),
\end{equation}
where $\Delta$ is some diagonal matrix with $\alpha=d(\Delta)$,
$\alpha\geq 0$ and $\sum_{i=1}^{r_{\max}}\alpha_i =1$. In the case
where the multiplicity at the optimum is one, the optimal
eigenbasis reduces to a unique vector and we saw in Proposition
\ref{multone} that multiplying this vector by $\sqrt{n+1}$ gives a binary
solution. Now let us turn to the case where there are $r_{\max}>1$
eigenvectors. To each unit norm eigenvector $e^j$, we associate a
subgradient $g_j=[(n+1)(e^j_1)^2-1, \ldots,(n+1)(e_{n+1}^j)^2-1]^t$. Then,
(\ref{eigopt}) implies that
\begin{equation}
\label{convcombzero} 0=\sum_{j=1}^{r_{\max}} \alpha_j g_j.
\nonumber
\end{equation}
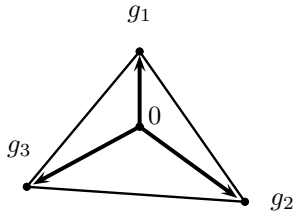
\begin{figure}[htb]
\label{opteig}
\begin{center}
\begin{pspicture}(-1,-1)(2,2)
\psset{showpoints=true, 
       dotstyle=*, 
       dotsize=3pt, 
       linewidth=1.5pt, 
       subgriddiv=1,       
       gridcolor=lightgray} 
\psline{->}(0,0)(0,.95)
\psline{->}(0,0)(1.3,-.95)
\psline{->}(0,0)(-1.42,-.77)
\psset{linewidth=.9pt} 
\psline{-}(0,1)(1.4,-1)
\psline{-}(1.4,-1)(-1.5,-.8)
\psline{-}(0,1)(-1.5,-.8)
\psset{linewidth=0.5pt} 
\rput{0}(0,1.5){$g_1$}
\rput{0}(1.9,-1){$g_2$}
\rput{0}(-1.6,-.3){$g_3$}
\rput{0}(.2,.15){$0$}
\end{pspicture}
\end{center}
\caption{Three subgradients in $\mathbb R^2$ at the optimal dual solution, one convex combination of which gives zero.}
\end{figure}
Now one natural strategy might be the following: pick the best
eigenvector, i.e. the eigenvector $\sqrt{n+1}e^{j_0}$ whose
associated coefficient $\alpha_{j_0}$ in expression
(\ref{convcombzero}) is the {\em greatest} and round its
coordinates to the nearest binary values. There is a second
strategy : draw random linear combinations of the
$\sqrt{n+1}e^j$'s giving preference to the components with higher
associated coefficient in (\ref{convcombzero}). This can be done
by sampling vectors of the type
$$
\sum_{j=1}^{r_{\max}} \zeta_j \sqrt{n+1}e^j
$$
where the $\zeta_j$'s are independent random variables with
distribution $\mathcal N(0,\alpha_j)$. For each sample, a feasible
solution is obtained by rounding off the components to the nearest
binary. We sum up this procedure as follows.
\begin{proc}{\bf (Randomized algorithm based on optimal eigenvectors)}
\label{eigGW}
1. Find the matrix $E_{\max}$ whose columns form an orthonormal
eigenbasis associated to $\lambda_{\max} (A(u^*))$ such that
(\ref{convcombzero}) holds for some $\alpha_j$'s satisfying
$\alpha\geq 0$ and $\sum_{j=1}^{r_{\max}}\alpha_j=1$.

2. Let $\zeta$ be a random vector with distribution $\mathcal
N(0,D(\alpha))$. The random cut is defined by
$$
Z={\rm sign} \Big( \sqrt{n+1}E_{\max}\zeta \Big).
$$

3. Draw $n$ samples from $Z$, say $Z^1$, \ldots, $Z^n$ and choose
the sample giving the best value of the objective function $z^tMz$.
\end{proc}
The important result is that this second strategy is equivalent to
Goemans and Williamson's randomized procedure.

\begin{prop}
Procedure \ref{eigGW} is equivalent to Goemans and Williamson's
algorithm.
\end{prop}
{\bf Proof}. 
Set $W=E_{\max}D(\alpha)^{\frac12}$. Then Theorem \ref{equiv} and equation
\ref{convcombzero} imply that $X^*=V^tV$ with $V^t=W$, thus retrieving the
Cholesky factorization of $X^*$. Let $\xi=D(\alpha)^{-\frac12}\zeta$. It
is clear that $\xi$ has distribution $\mathcal N(0,I)$. This
proves that the cut $Z$ obtained by Procedure
\ref{eigGW} is exactly the output of Goemans and Williamson's
procedure.
\hfill $\Box$

The eigenvalue point of view thus allowed us to provide an alternative and geometric
explanation for taking a random cut using a uniformly distributed
variable on the sphere in  Goemans and Williamson's methodology.

\section{Two application examples}
In this section, we provide some results for the concrete problems of image denoising and show how this 
relaxation applies to the problem of multiuser detection in CDMA systems. 

\subsection{Image denoising}
\subsubsection{Presentation of the problem}
The first set of simulations is devoted to the denoising problem,
in which $A$ is simply the identity matrix. This is the problem
considered in \cite{ICIP98:Nikolova}, \cite{Biometrika00:Gibbs} and \cite{JRSS86:Besag} for
instance. The original binary image as 26 rows and 62 columns
which gives a total number of 1612 variables. 

For this problem, the penalization
matrix $P$ is chosen so as to smooth the image. This is achieved
by requiring neighboring pixels to be similar in the sense that if
$i$ and $j$ are indices of neighbor pixels, then, we would like
the least square cost to be penalized by the quantity
$|x_i-x_j|^2$. Thus, $P$ is the matrix associated to the quadratic
form
\begin{equation}
\label{reg}
\sum_{i\sim j} \zeta_{ij}|x_i-x_j|^2,
\end{equation}
where $i\sim j$ denotes the property of being neighbor indices and the $\zeta_{ij}$ are nonnegative.
The neighborhood  of each pixel is usually chosen to be the north, south, east and west pixels.

\subsubsection{Exactness of the relaxation}
The following theorem is the main result of this section. 
\begin{thm} For $A=I$, the identity matrix and $P$ the matrix associated to the quadratic form (\ref{reg}), the eigenvalue relaxation 
is exact.  
\end{thm}
{\bf Proof}. The eigenvalue relaxation of the optimization problem corresponding to this binary least square denoising problem is as 
before 
\begin{equation}
\min_{u\in \mathbb R^{n+1}}(n+1) \lambda_{\max}(-(M+\nu P+\frac{e^t u}{n+1}I)+D(u_1^n)). \hspace{1cm} (Denoise)
\nonumber
\end{equation}
Consider now the perturbed optimization problem 
\begin{equation}
\min_{u\in \mathbb R^{n+1}} (n+1) \lambda_{\max}(-(M+\Delta M+\frac{e^t u}{n+1}I)+D(u_1^n)+) \hspace{1cm} (Perturbed) 
\nonumber
\end{equation}
where $\Delta M$ is negative outside the diagonal. Since the $\zeta_{ij}$ are nonnegative, the matrix $P$ has only 
nonpositive off diagonal terms and thus, Theorem \ref{strong} proves that strong duality holds for this problem and there exists 
a binary eigenvector that achieves optimality. Assume that $\Delta M$ is chosen so that $\|\Delta M\|\leq \epsilon$. Then, the optimum value $\theta^*$
of problem (Denoise) and the optimum value $\theta^*_{\Delta M}$ of problem (Perturbed) satisfy 
\begin{equation}
\theta^*_{\Delta M}-(n+1)\epsilon \leq \theta^* \leq \theta^*_{\Delta M}+(n+1)\epsilon.
\nonumber
\end{equation}
Moreover, by weak duality, we have 
\begin{equation}
\max_{x\in \{-1,1\}^n} -x^t (I+\nu P)x \leq \theta^*_{\Delta M}.
\nonumber
\end{equation}
Since strong duality holds for problem (Perturbed), denoting by $x^*_{\Delta M}$ a solution of $\max_{x\in \{-1,1\}^n} -x^t (I+\Delta M+\nu P)x$ we have 
\begin{equation}
\theta^*_{\Delta M}= -{x^*_{\Delta M}}^t (I+\Delta M+\nu P) x^*_{\Delta M}\leq \max_{x\in \{-1,1\}^n} -x^t (I+\nu P)x.
\nonumber
\end{equation}
Therefore, we obtain 
\begin{equation}
-{x^*_{\Delta M}}^t (I+\Delta M+\nu P) x^*_{\Delta M}\leq \max_{x\in \{-1,1\}^n} -x^t (I+\nu P)x \leq -{x^*_{\Delta M}}^t (I+\Delta M+\nu P) x^*_{\Delta M}+(n+1)\epsilon,
\nonumber
\end{equation}
which implies 
\begin{equation}
\label{presque}
-{x^*_{\Delta M}}^t (I+\nu P) x^*_{\Delta M}-(n+1)\epsilon
\leq \max_{x\in \{-1,1\}^n} -x^t (I+\nu P)x \leq -{x^*_{\Delta M}}^t (I+\nu P) x^*_{\Delta M}+2(n+1)\epsilon,
\nonumber
\end{equation}
Now, since $\{-1,1\}^n$ is finite, the image $\mathcal I$ of $\{-1,1\}^n$ by the function $-x^t (I+\nu P)x$ is a finite set. Let $\delta$ denote the closest number to 
$\max_{x\in \{-1,1\}^n} -x^t (I+\nu P)x$ in $\mathcal I$. Now, choosing $2(n+1)\epsilon<\delta$, we obtain 
\begin{equation}
\label{presque2}
-{x^*_{\Delta M}}^t (I+\nu P) x^*_{\Delta M}=\max_{x\in \{-1,1\}^n} -x^t (I+\nu P)x
\nonumber
\end{equation}    
which proves that the denoising problem is polynomial time solvable by solving problem (Perturbed).  \hfill$\Box$

This theorem is to be compared with the results of D. M. Greig, B. T. Porteous and A. H. Seheult \cite{Greig:JRSSB} which formulates 
the binary denoising problem as a minimization problem with cost given at the top of page 273. The objective to be minimized 
in \cite{Greig:JRSSB} can be rearranged so as to minimize a linear cost with same penalization as the one given by 
(\ref{reg}). The main contribution of \cite{Greig:JRSSB} is to say that this problem can be solved in polynomial time using a network flow algorithm. 
Notice that our proof works for $A^tA=0$ and any additional linear term added to the penalized objective function to be optimized. Since 
the eigenvalue relaxation can also be optimized in polynomial time,  
this confirms that the eigenvalue relaxation performs at least as good as previous approaches on a well known problem. 
On the other hand, the eigenvalue relaxation can be a flexible approach in more complicated cases where $A$ is not equal to the identify or 
other quadratic constraints have to be incorporated such as in \cite{IEEEPAMI03:Keuchel}. 

\subsubsection{A numerical experiment}
The experiments reported on below were performed for the case of quite noisy original images. 
The noise was taken to be additive, independent identically distributed and Gaussian $\mathcal N(0,2)$
and was applied to the symmetrized image with pixel values in $\{-1,1\}$. 
In order to show the influence of
the smoothing parameter $\nu$, we displayed the percentage of misspecified bits vs values of $\nu$. 
The recovered image is the one with the choice of $\nu$ giving the best percentage of bits recovered.  

We found the results very encouraging. Indeed, even
when the observed image is very noisy, we still recover an image
which is readable. This suggested that an appropriate
postprocessing might easily allow to recover the original written
words, by comparing the letters to a given dictionary.
Cross validation can be used to estimate $\nu$. We will not discuss this problem here. Instead, it
seems reasonable to argue that the choice of $\nu$ can just be
made {\em a posteriori} since it consists of tuning the
method until a satisfactory solution is obtained. This reduces the
hard combinatorial initial problem to a simpler one parameter
knobing procedure. The displayed experiment and the numerous
simulations not presented here confirm that robust intervals for
the values of $\nu$ are not very difficult to identify in
practice.



\subsection{Multiuser detection in CDMA systems}
\subsubsection{Presentation of the problem}
This problem was studied by \cite{IT86:Verdu} using the maximum likelihood approach. As we will see, 
the resulting optimization problem is of the same form as the binary least squares problem. The main difference here 
is that $A\neq I$ and $P=0$. 
 
A synchronous K users DS-CDMA system is considered with a common single path additive white Gaussian noise (AWGN) channel.      
The signature waveform of the $k$th user is denoted by $s_k(t)$, a function taking nonzero values in $[0,T]$ 
and being equal to zero outside this interval, and $x_k$ is the information bit transmitted by user $k$. The 
overall received signal is therefore of the form 
$$
y(t)=\sum_{k=1}^K a_k x_k s_k(t) + n(t) 
$$ 
where $a_k$ is the amplitude of the $k$th user's signal and $n(t)$ is an additive white Gaussian white noise with 
zero mean and variance $\sigma^2$. The signal $y$ is then filtered using a bank of $K$ matched filters. The output of the 
$k$th matched filter is given by 
$$
y_k=\int_{0}^{T} y(t)s_k(t) dt. 
$$
In matrix form, this can be written
$$
y=RA x + \nu
$$
where $y=[y_1,\ldots,y_k]^t$, $R$ is the correlation matrix whose components are given by 
$R_{ij}=\int_{0}^{T} s_i(t)s_j(t) dt$, $A=D(a)$ and $\nu$ is the vector with components 
$\nu_k=\int_{0}^{T} n(t)s_j(t) dt$. 

Since the gaussian vector has a correlation matrix equal to $\sigma^2 R$, the ML estimator is obtained
by simply solving the following combinatorial optimization problem. 
\begin{equation}
\label{ml}
\begin{array}{c}
\min_{x\in \mathbb R^n} x^t ARA x -2y^t A x \\
\\
\text{ s.t. } x_i\in \{-1,1\}, \hspace{.3cm} i=1,\ldots,K. 
\end{array}
\end{equation}

\subsubsection{Some comments}
The SDP approach seems to have been first applied for the DS-CDMA detection problem in \cite{SelectAreaCom01:Tan}. Since then 
numerous contributions have appeared using the SDR and comparing it to other methods as in \cite{Com04:Hasegawa} and 
\cite{Wireless04:Tan}. Extension to M-ary phase shift keying symbol constellations is proposed in \cite{SP04:Ma}. 
The issue of accelerating the speed of the method is addressed in \cite{SPLetters02:Abdi}. 
However, as for the former problem, the main drawback of the standard primal semidefinite relaxation is that the size of 
the problem is greatly increased by using $K\times K$ matrices instead of vectors of size $K$. 
In order to overcome this problem, a better approach using semidefinite programming duality was recently proposed in \cite{SP03:Wang}. 

The analysis of the previous sections proves that the eigenvalue relaxation is equally applicable to this problem and maybe 
a good competitor to the SDP relaxation. The most important point of our analysis is the following: Theorem \ref{strong} proves that 
if the correlation matrix $R$ is componentwise negative outside the diagonal, then strong duality holds, i.e. the detection problem
can be solved exactly in polynomial time. The construction of efficient signatures is the current subject of an active research activity. 
For instance, the theory of frames allows to consider the problem from an interesting viewpoint as developed in \cite{Tropp:IEEEIT05}.
Our findings suggest in particular that the componentwise negativity of the correlation matrix may be an interesting constraint 
to look at in future investigations on this problem. 

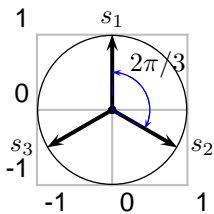
\begin{figure}[htb]
\label{negcorr}
\begin{center}
\begin{pspicture}(-1,-1)(1,1)
\psset{showpoints=true, 
       dotstyle=*, 
       dotsize=3pt, 
       linewidth=1.5pt, 
       subgriddiv=1,       
       gridcolor=lightgray} 

\psgrid
\psline{->}(0,0)(0,1)
\psline{->}(0,0)(.866,-.5)
\psline{->}(0,0)(-.866,-.5)
\psset{linewidth=0.5pt} 
\pscircle(0,0){1}
\psarc[linecolor=blue]{<->}(0,0){.5}{-30}{90}
\rput{0}(0,1.2){$s_1$}
\rput{0}(1.2,-.5){$s_2$}
\rput{0}(-1.2,-.5){$s_3$}
\rput{0}(.6,.6){$2\pi/3$}
\end{pspicture}
\end{center}
\caption{Three vectors in $\mathbb R^2$ with correlation matrix having negative off-diagonal components.}
\end{figure}

Finally, the eigenvalue relaxation can also be useful even for general signatures because of the weak duality property. Indeed, 
several recent publications prove that clever heuristics can perform better than the SDP relaxation. However, in real situations it 
is hard to certify that a primal solution provided by such a heuristic is indeed the optimal solution because the original signal 
is unknown. Comparing the dual optimal value to a primal value given by a heuristic can give a precise idea of the error without
prior information on the signal. 

\subsubsection{A numerical experiment}
\begin{center}
\begin{figure}[htb]
\label{cdma}
\includegraphics[width=16cm]{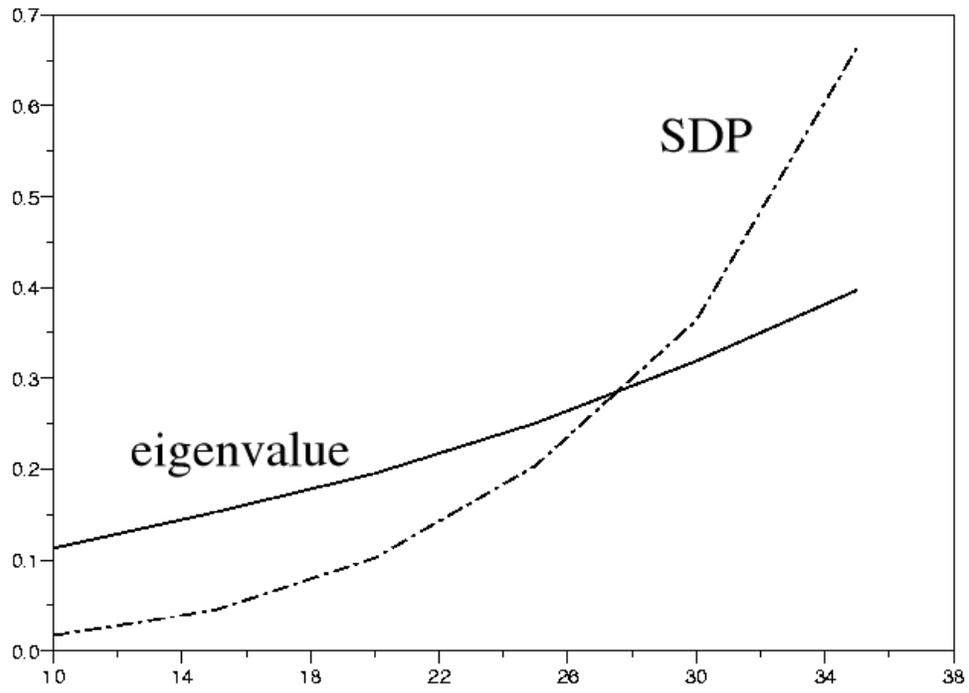}
\caption{Comparison of SDP and eigenvalue relaxations for CDMA multiuser detection}
\end{figure}
\end{center}

In order to verify this point, we performed Monte Carlo simulations over 1000 random problems for a number a users varying from 10 to 35. These 
computational experiments are reported in Figure \ref{cdma} where the number of users is on the x-axis and the 
average computation time is on the y-axis. The computations where performed using the Scilab software \cite{Web:Scilab}. 
The SDP solver called {\em Semidef} interfaces Boyd and Vandenberghe's sp.c program. The eigenvalue relaxation was 
solved using the solver {\em Optim} with the "nd" option for possibly nondifferentiable costs as is the case here. 
The curves in Figure \ref{cdma} interpolate the average computation times for messages taken to be sequences of 
 uniform and independant variables taking values in $\{0,1\}$ vs. the number of users. The curve with dashed style is 
for the results of the SDP relaxation while the curve with plain style is for the eigenvalue relaxation. 
Our computations suggest that the eigenvalue relaxation has lower complexity growth as the number of users increases exactly as expected. The reader should 
be warned that this experiment does not prove that the complexity of the eigenvalue relaxation is lower than the SDP relaxation. The experiment 
only shows that when a widely used routine for SDP is used, the eigenvalue relaxation, solved using a general purpose bundle method 
available through a free a well established software, has a lower complexity growth on this problem.

\section{Appendix: Arrow matrices and strict interlacing of eigenvalues}
\label{app}
Arrow matrices are matrices $A$ of the form 
$$
A= \left[
\begin{array}{cc}
D(a) & b \\
b^t & c
\end{array}
\right],
$$
The properties of the eigenvalues of such matrices have been well studied in the past. Some of them are
summarized in the following theorem.  
{\bf Theorem A}. Let $A$ be an arrow matrix, with $a_1\leq a_2
\leq \ldots \leq a_n$. Moreover, assume that all the components of $b$ are different
from zero. Let $\lambda_1\leq \lambda_2 \leq ... \leq
\lambda_{n+1}$ be its eigenvalues considered in increasing order.
Then, the characteristic polynomial of $A$ is given by
$$
p_A(\lambda)=(c-\lambda) \prod_{i=1}^n (a_i-\lambda)-\sum_{i=1}^n
\prod_{j\neq i} (a_j-\lambda)b_i^2.
$$
Then, we have $\lambda_1<a_1$ and $a_n<\lambda_{n+1}$. Moreover,if
$a_i=a_{i+1}$ we have $a_i=\lambda_{i+1}=a_{i+1}$ and if
$a_i<a_{i+1}$, we have $a_i<\lambda_{i+1}<a_{i+1}$.

The properties of the eigenvalues of arrow matrices are part of the folkore, especially in the 
realm of mathematical physics. We give a sketch of 
the proof of this theorm below in order to give the main ideas underlying the results. 

{\bf Proof of Theorem A}. 
The formula for the characteristic polynomial $p_A(\lambda)={\rm det}(A-\lambda
I)$ is easily obtained by reccurence on the dimension. 
We have to consider two cases: 
\begin{itemize}
\item for some $i$, $a_i=a_{i+1}$,
\item $a_1<a_2<\ldots<a_n$ 
\end{itemize}  
In the first case $a_i$ is a root of $p_A$. In the second case
$p_A(a_i)=\prod_{j\neq i}(a_j-a_i) b_i^2$ which is different from 
zero since we assumed all the $b_i$'s to be different from zero.
In this case, the eigenvalues of $A$ are the zeros of the function
$$
q_A(\lambda)=c-\lambda + \sum_{i=1}^n \frac{b_i^2}{\lambda-a_i}.
$$
From this formula, we deduce that there is a root in each interval
$(-\infty,a_1)$, $(a_i,a_{i+1})$, for all $i=1,\ldots,n$ and
$(a_n,+\infty)$. 

The final conclusions are easily derived by combining the results in the 
two simple cases discussed above. 
\hfill$\Box$ 

\section{Conclusion}

In this paper, we surveyed the main properties of the eigenvalue relaxation for
binary least squares problem. A full connection with the standard SDP relaxation was presented 
and we showed how to recover a solution of the Semi-Definite program from the solution of the eigenvalue 
minimization problem. The problem of recovering primal binary solution was also addressed and we 
gave simple sufficient conditions for strong duality. In the case where these conditions are not satisfied, 
the randomized procedure adapted from Goemans and Williamson's allows to recover binary solutions with garanteed relative approximation ratio
due to Nesterov's bound. Two applications were presented: binary image denoising and detection in multiuser CDMA systems. In the 
case of image denoising, we show that strong duality holds. For the multiuser detection problem, our results prove that 
strong duality holds when the signature covariance matrix has nonpositive off diagonal components.

\newpage
\begin{center}
\begin{figure}[htb]
\includegraphics[width=16cm]{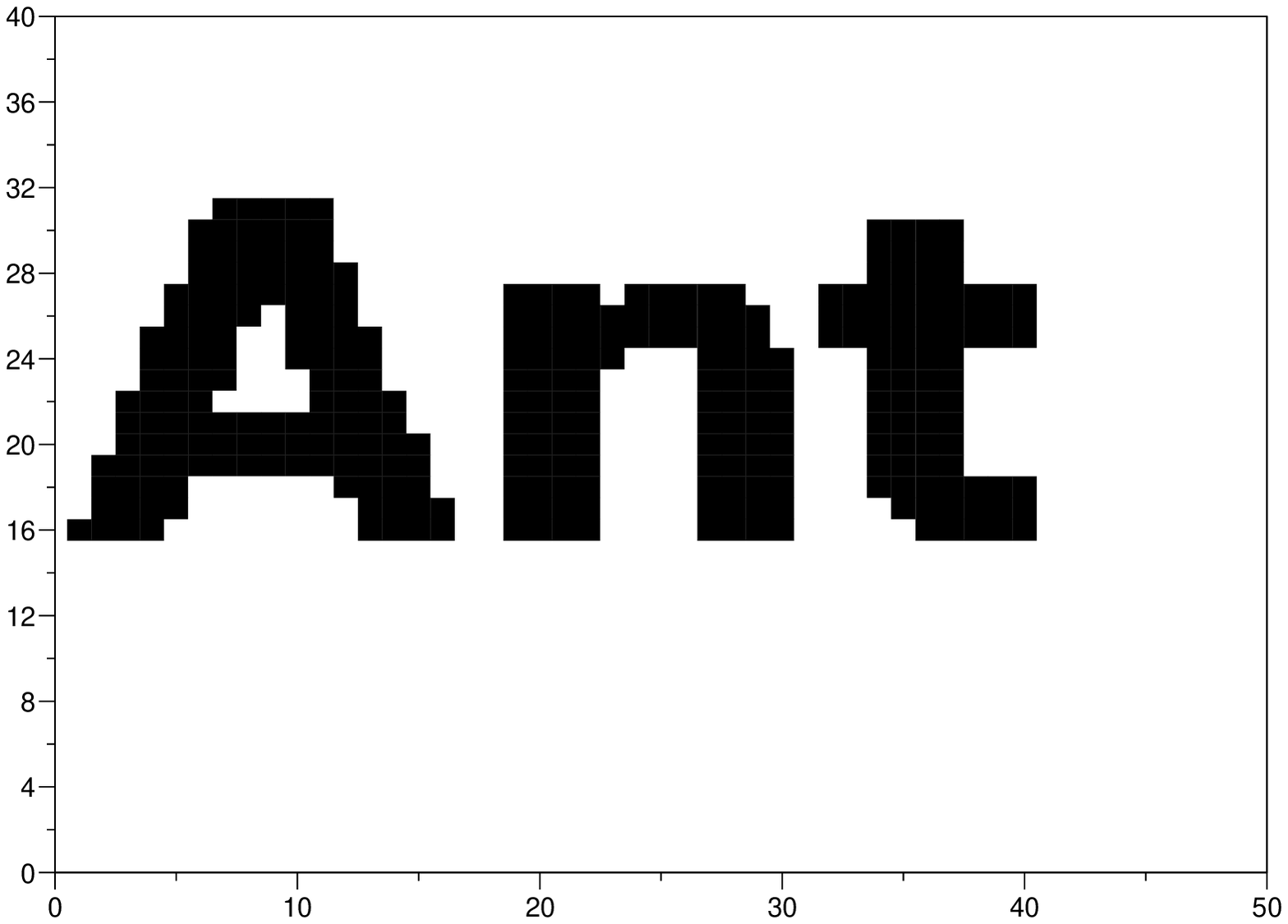} 
\caption{Original image}
\end{figure}
\end{center}

\newpage

\begin{center}
\begin{figure}[htb]
\includegraphics[width=16cm]{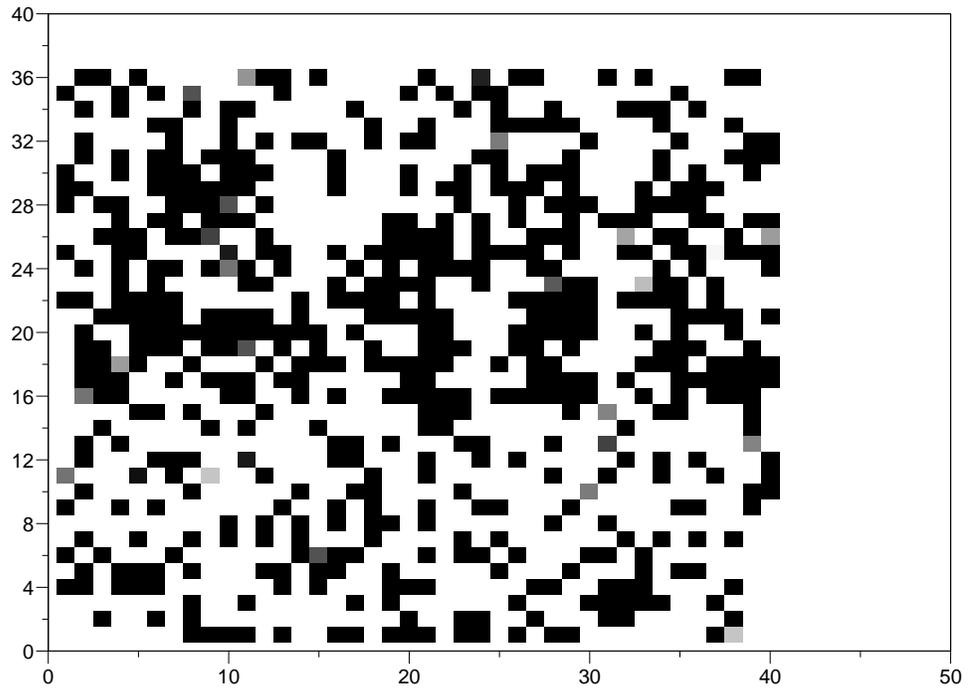} 
\caption{Noisy image: i.i.d. $\mathcal N(0,2)$}
\end{figure}
\end{center}

\newpage

\begin{center}
\begin{figure}[htb]
\includegraphics[width=16cm]{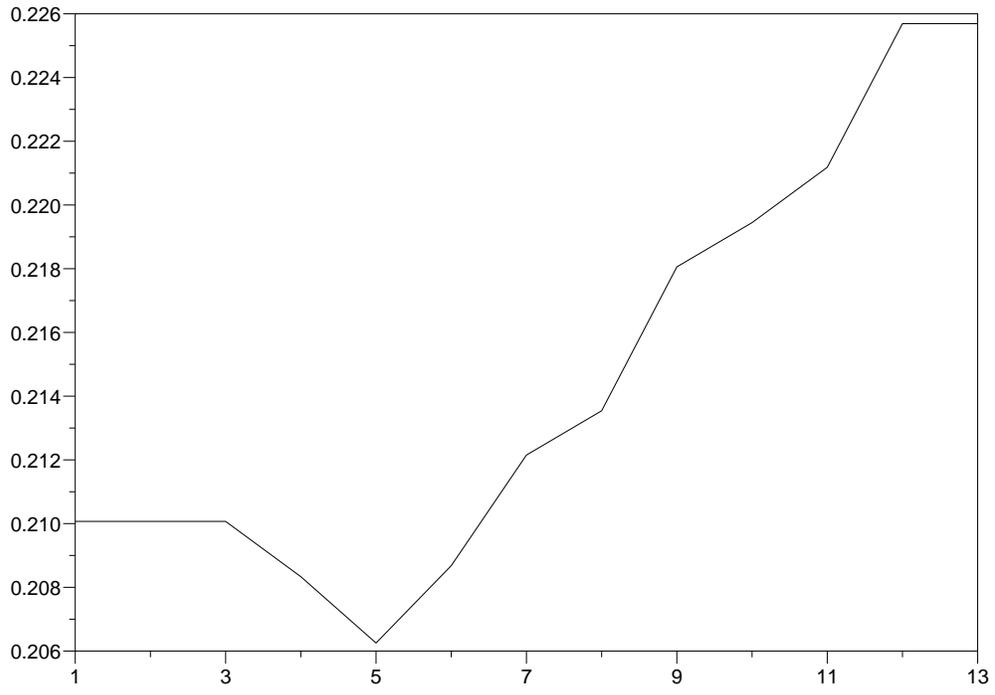}
\caption{Percentage of misspecified bits v.s. $\nu$}
\end{figure}
\end{center}

\newpage

\begin{center}
\begin{figure}[htb]
\includegraphics[width=16cm]{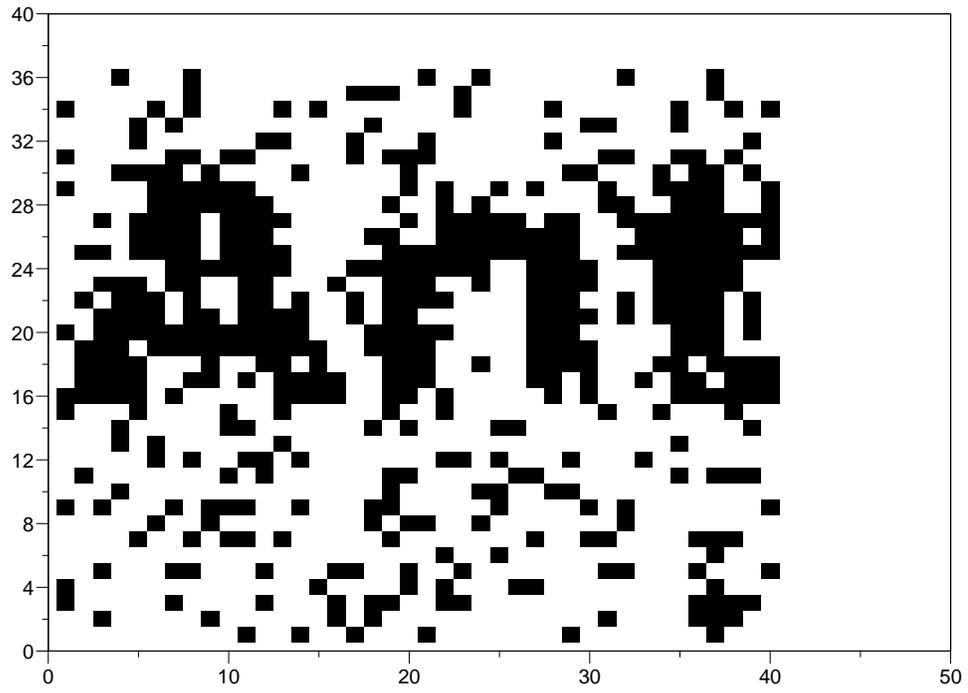}
\caption{Recovered image}
\end{figure}
\end{center}

\newpage

\end{document}